\documentstyle[preprint,aps]{revtex}

\begin{document}
\title{Manipulating the Speed of Sound in a Two-Component Bose-Einstein Condensate}
\author{C. P. Search and P.R. Berman}
\address{Michigan Center for Theoretical Physics, Physics Department, University\\
of Michigan, Ann Arbor, MI, 48109-1120}
\date{\today }
\maketitle

\begin{abstract}
We consider a two-component weakly interacting Bose-Einstein condensate in
the presence of an external field which couples the two components. We
express the Hamiltonian in terms of the energy eigenstates of the
single-body part of the Hamiltonian. These eigenstates are the atomic
dressed states of quantum optics. When the energy difference between the two
dressed states is much larger than the mean-field interactions, two-body
interactions in the dressed state basis that do not conserve the number of
atoms in each of the two dressed states are highly suppressed. The two-body
interactions then take on a simplified form in the dressed basis with
effective coupling constants that depend on the intensity and frequency of
the external field. This implies that the chemical potential as well as the
quasiparticle spectrum may be controlled experimentally in a simple manner.
We demonstrate this by showing that one may achieve significant variations
in the speed of sound in the condensate, a quantity which has been measured
experimentally.
\end{abstract}

\section{Introduction}

The recent experimental realization of trapped Bose-Einstein condensates
(BEC) with internal degrees of freedom corresponding to different hyperfine
states \cite{Na}\cite{Rb} has sparked much theoretical and experimental
study of the properties of multi-component condensates. Multi-component
condensates exhibit a rich variety of new phenomena not present in
condensates with a scalar order parameter. For condensates with a vectorial
order parameter, such as the $F=1$ hyperfine multiplet of $^{23}Na$, there
have been predictions of spin waves, instability of vortices with more than
one unit of circulation, coreless vortices, dynamic spin localization \cite
{Na2} \cite{Na3}\cite{Na4} and the observation of spin domains\cite{spin}.
For two-component condensates, such as $^{87}Rb$, there have been extensive
theoretical and experimental investigations of the dynamics of the
condensates in the presence of external fields which couple the components.
Examples include Ramsey fringes\cite{ramsey}, non-linear Josephson
oscillations \cite{Joseph}, instabilities in the quasiparticle spectrum \cite
{instable}, the elimination of the mean-field shift in the energy \cite
{shift}, control of the spatial dependence of the two components \cite
{dressed}, and collapse and revival of Rabi oscillations \cite{collapse}.

In this paper, a two-component homogenous weakly interacting BEC at zero
temperature in the presence of an external field which couples the two
states is investigated. The physical system is similar to that discussed in 
\cite{instable}. In quantum optics, the dressed states, which are the energy
eigenstates of a two level atom interacting with an external radiation
field, form a convenient basis for many problems \cite{dressedatom}. More
specifically, if one has a two-level system with states $\left|
a\right\rangle $ and $\left| b\right\rangle $ governed by the Hamiltonian, 
\[
H=\frac{\hbar }{2}\left( 
\begin{array}{cc}
\delta  & 2\Omega _{R} \\ 
2\Omega _{R} & -\delta 
\end{array}
\right) ,
\]
then the dressed states, $\left| c\right\rangle $ and $\left| d\right\rangle 
$, are simply the energy eigenstates of $H$ with eigenvalues $\pm \frac{%
\hbar }{2}\omega _{cd}$. The dressed states are a superposition of the
states $\left| a\right\rangle $ and $\left| b\right\rangle $ with amplitudes
determined by $\delta /\Omega _{R}$. We explore the use of a dressed state
basis for describing an interacting two-component condensate.

When the two-body interactions in the second-quantized Hamiltonian are
rewritten in terms of a dressed state basis, one finds that there are terms
that conserve the number of atoms in each of the two dressed states and
terms that change the number of atoms in each of the dressed states. When
the energy difference between the dressed states is much larger than the
single particle kinetic energies and the mean-field energy, the Hamiltonian
simplifies since the terms that change the number of atoms in each of the
dressed states may be neglected. In this limit, the ground state of the
condensate consists of atoms in one of the dressed states only, and the
calculation of the excited states becomes trivial. In this case, there are
two branches to the spectrum of elementary excitations. One branch has the
standard Bogoliubov dispersion relation while the other branch corresponds
to single particle excitations. In addition, the speed of sound is an
explicit function of $\delta /\Omega _{R}$. This result indicates that the
spectrum of collective excitations of the condensate depend on the coherence
between the internal states of the atoms in a manner which can be
experimentally controlled.

Before proceeding, it is helpful to review the hydrodynamics of a single
component BEC at zero temperature. At zero temperature, the thermal
component of the Bose gas is absent and the quantum depletion is negligible
if the bosons are weakly interacting. In this case, the wave function (or
order parameter) for the condensate, $\phi ({\bf r},t)=\sqrt{n({\bf r},t)}%
e^{iS({\bf r},t)}$, obeys the time-dependent Gross-Pitaevskii equation \cite
{Gross} which may be written as two coupled equations for the density, $n(%
{\bf r},t),$ and velocity, ${\bf v}({\bf r},t)=\frac{\hbar }{m}\nabla S({\bf %
r},t)$ \cite{stringari}, 
\begin{equation}
\frac{\partial }{\partial t}n+\nabla \cdot \left( {\bf v}n\right) =0
\label{2}
\end{equation}
and 
\begin{equation}
m\frac{\partial }{\partial t}{\bf v}+\nabla \left( V({\bf r})+Un-\frac{\hbar
^{2}}{2m\sqrt{n}}\nabla ^{2}\sqrt{n}+\frac{m|{\bf v|}^{2}}{2}\right) =0.
\label{3}
\end{equation}
Here $U=\frac{4\pi \hbar ^{2}a}{m}$, $a$ is the s-wave scattering length
which characterizes the two-body interactions between atoms, $m$ is the
atomic mass, and $V({\bf r})$ is an external potential. In general, the
quantum pressure, $\frac{\hbar ^{2}}{2m\sqrt{n}}\nabla ^{2}\sqrt{n},$ may be
neglected in comparison to the mean-field interaction. For a static
condensate ground state, this corresponds to the Thomas-Fermi limit for the
density, 
\begin{equation}
n_{o}({\bf r})=U^{-1}(\mu -V({\bf r})),  \label{TF}
\end{equation}
where $\mu $ is the chemical potential. By considering small fluctuations, $%
\delta n({\bf r},t),$ about the static ground state density $n_{o}({\bf r})$%
, one may derive a linear wave equation for $\delta n({\bf r},t)$ from Eqs. (%
\ref{2}-\ref{3}), 
\begin{equation}
\partial _{t}^{2}\delta n({\bf r},t)=\nabla \cdot \left[ u^{2}({\bf r}%
)\nabla \delta n({\bf r},t)\right] ,  \label{wave}
\end{equation}
where $u^{2}({\bf r})=\frac{U}{m}n_{o}({\bf r})$ is the local speed of
sound. The sound speed may also be obtained from the relation, $u^{2}({\bf r}%
)=\frac{1}{m}\frac{\partial P}{\partial n}$, where $P$ is the pressure of
the ground state of the condensate. The derivation of Eq. (\ref{wave})
neglects the quantum pressure term in Eq. (\ref{3}). When $V({\bf r})\equiv 0
$, the condensate ground state is spatially homogenous so that $n_{o}({\bf r}%
)=n_{o}$. In this case, the qauntum pressure may be neglected when the
variations in the density are small over distances on the order of the
healing length, $\xi =1/\sqrt{8\pi an_{o}},$ or equivalently, the elementary
excitations have momenta satisfying $\hbar k\ll mu$ where $k$ is the wave
vector for the excitation and $u=u({\bf r})$.

For a homogenous BEC (i.e. $V({\bf r})\equiv 0$), the energies of the
elementary excitations form a continuous spectrum. The dispersion relation
for the long wavelength collective excitations of the condensate have the
form $\omega =uk\ll \mu /\hbar $. These excitations correspond to phonons.
The speed of sound in a homogenous BEC (sometimes called Bogoliubov or zero
sound) was first derived by Bogoliubov \cite{bogo} using a microscopic
theory of weakly interacting bosons, 
\begin{equation}
u=\sqrt{\frac{4\pi \hbar ^{2}an_{o}}{m^{2}}}.  \label{sound}
\end{equation}
The Bogoliubov theory is used in this paper to derive the excitation spectra
for a two component condensate.

Bogoliubov sound in a condensate of sodium atoms has been experimentally
studied by M. R. Andrews and coworkers in a highly elongated cigar shaped
trap\cite{sound}. Note that $V({\bf r})\neq 0$ in this case but along the
long axis of the trap, the condensate can be treated as being locally
homogenous. In this experiment, the repulsive dipole force of a focused
blue-detuned laser beam was used to create a localized density perturbation
at the center of the trap. The double peaked perturbation was subsequently
imaged as it propagated along the long axis of the trap. When the density of
atoms in the trap was varied, the measured sound speeds showed good
agreement with Eq. (\ref{sound}) where $n_{o}$ refers to the density at the
center of the trap divided by $2$. The extra factor of a $1/2$ has a simple
interpretation. Since the sound propogation is confined to the long axis of
the trap, the condensate can be treated as an effective one-dimensional
system. In this case the effective density is given by averaging the
condensate density over the directions transverse to the long axis which in
the Thomas-Fermi limit is just the central density divided by $2$ \cite
{soundtheory}.

This experiment has been analyzed theoretically by several authors \cite
{soundtheory}\cite{stringari2} in the Thomas-Fermi limit. In a trapped
condensate, sound waves can propagate provided that the excitations satisfy
both $\hbar k\ll mu({\bf r})$ and $kL\gg 1$ where $L$ is the size of the
condensate in the direction of propagation. The latter condition corresponds
to wavelengths much less than the condensate size so that the variations in $%
n_{o}({\bf r})$ over distances on the order of the wavelength are
negligible. Consequently, the condensate can be treated as locally
homogenous. As such, the results derived for a homogenous BEC using
Bogoliubov theory may be used for trapped condensates with the substitution $%
n_{o}\rightarrow \bar{n}_{o}({\bf r})$ where the bar denotes the possible
averaging over transverse dimensions for anisotropic condensates.

In the following section a second-quantized Hamiltonian for a two-component
BEC is given in the ''bare'' atomic basis and the Hamiltonian in the dressed
basis is derived. In section III, the condensate ground state and elementary
excitations in the dressed basis are determined. Finally, in section IV, the
results are discussed including justification of the physical approximations
used.

\section{Physical Model}

We consider a collection of $N$ bosonic atoms that have internal states $%
\left| a\right\rangle $ and $\left| b\right\rangle $ with energies $\hbar
\omega _{o}/2$ and $-\hbar \omega _{o}/2$, respectively. There is a
spatially uniform radiation field with frequency $\omega _{e}$ which couples
the two internal states with a Rabi frequency $\Omega _{R}$ \cite{two-photon}%
. The atom field detuning is denoted by $\delta =\omega _{o}-\omega _{e}$.
The atoms in states $\left| a\right\rangle $ and $\left| b\right\rangle $
may also be subject to an external trapping potential $V_{a}({\bf r)}$ and $%
V_{b}({\bf r)}$, respectively. Furthermore, the atoms interact via elastic
two-body collisions through the interaction potentials $V_{ij}({\bf r}-{\bf r%
}^{\prime })$ for $i,j=\{a,b\}$. The many-body Hamiltonian operator
describing the system is given by, 
\begin{mathletters}
\begin{eqnarray}
\hat{H} &=&\hat{H}_{atom}+\hat{H}_{coll}  \label{ham} \\
\hat{H}_{atom} &=&\int d^{3}r\left\{ \hat{\Psi}_{a}^{\dagger }({\bf r})\left[
-\frac{\hbar ^{2}}{2m}\nabla ^{2}+V_{a}({\bf r)+}\frac{\hbar \delta }{2}%
\right] \hat{\Psi}_{a}({\bf r})+\hat{\Psi}_{b}^{\dagger }({\bf r})\left[ -%
\frac{\hbar ^{2}}{2m}\nabla ^{2}+V_{b}({\bf r)-}\frac{\hbar \delta }{2}%
\right] \hat{\Psi}_{b}({\bf r})\right.  \nonumber \\
&&\left. +\hbar \Omega _{R}\left[ \hat{\Psi}_{a}^{\dagger }({\bf r})\hat{\Psi%
}_{b}({\bf r})+\hat{\Psi}_{b}^{\dagger }({\bf r})\hat{\Psi}_{a}({\bf r})%
\right] \right\}  \label{hatom} \\
\hat{H}_{coll} &=&\frac{1}{2}\int d^{3}rd^{3}r^{\prime }\left\{ \hat{\Psi}%
_{a}^{\dagger }({\bf r})\hat{\Psi}_{a}^{\dagger }({\bf r}^{\prime })V_{aa}(%
{\bf r}-{\bf r}^{\prime })\hat{\Psi}_{a}({\bf r}^{\prime })\hat{\Psi}_{a}(%
{\bf r})+\hat{\Psi}_{b}^{\dagger }({\bf r})\hat{\Psi}_{b}^{\dagger }({\bf r}%
^{\prime })V_{bb}({\bf r}-{\bf r}^{\prime })\hat{\Psi}_{b}({\bf r}^{\prime })%
\hat{\Psi}_{b}({\bf r})\right.  \nonumber \\
&&\left. +2\hat{\Psi}_{a}^{\dagger }({\bf r})\hat{\Psi}_{b}^{\dagger }({\bf r%
}^{\prime })V_{ab}({\bf r}-{\bf r}^{\prime })\hat{\Psi}_{a}({\bf r})\hat{\Psi%
}_{b}({\bf r}^{\prime })\right\} .  \label{hcoll}
\end{eqnarray}
Here, $\hat{H}_{atom}$ is the single particle Hamiltonian and $\hat{H}%
_{coll} $ represents two-body collisions.

The operators $\hat{\Psi}_{i}({\bf r})$ and $\hat{\Psi}_{i}^{\dagger }({\bf r%
})$ are bosonic annihilation and creation operators for an atom in state $%
i=\{a,b\}$ at position ${\bf r}$ which satisfy the commutation relations $[%
\hat{\Psi}_{i}({\bf r}),\hat{\Psi}_{j}^{\dagger }({\bf r}^{\prime })]=\delta
_{ij}\delta \left( {\bf r-r}^{\prime }\right) $ and $[\hat{\Psi}_{i}({\bf r}%
),\hat{\Psi}_{j}({\bf r}^{\prime })]=0$. The operators, $\hat{\Psi}_{i}({\bf %
r}),$ have been written in a field interaction representation which is
rotating at the frequency of the external field, $\omega _{e}$, 
\end{mathletters}
\begin{mathletters}
\begin{eqnarray}
\hat{\Psi}_{a}({\bf r}) &=&\hat{\Psi}_{a}^{(N)}({\bf r})e^{i\omega _{e}t/2};
\\
\hat{\Psi}_{b}({\bf r}) &=&\hat{\Psi}_{b}^{(N)}({\bf r})e^{-i\omega _{e}t/2};
\end{eqnarray}
where $\hat{\Psi}_{i}^{(N)}({\bf r})$ are the field operators in the normal
representation. This explains the appearance of the detuning in Eq. (\ref
{hatom}).

The two-body interaction between atoms, $V_{ij}({\bf r}-{\bf r}^{\prime }),$
depends on the internal states of the atoms. For a dilute gas such that $%
\bar{n}|a_{ij}|^{3}\ll 1$ where $\bar{n}$ is the average density and $a_{ij}$
is the s-wave scattering length between atoms in states $i$ and $j$, the
interaction may be written as $V_{ij}({\bf r}-{\bf r}^{\prime
})=U_{ij}\delta \left( {\bf r-r}^{\prime }\right) $ where $U_{ij}=\frac{4\pi
\hbar ^{2}a_{ij}}{m}$. It is assumed that $a_{ij}>0$ corresponding to
repulsive interactions.

For the remainder of this section and sections III and IV, we consider only
the case when $V_{i}({\bf r)}=0$. One can expand the field operators in a
basis of single particle momentum eigenstates, 
\end{mathletters}
\begin{mathletters}
\begin{eqnarray}
\hat{\Psi}_{a}({\bf r}) &=&\frac{1}{\sqrt{V}}\sum_{{\bf p}}\alpha _{a{\bf p}%
}e^{i{\bf p\cdot r}/\hbar }, \\
\hat{\Psi}_{b}({\bf r}) &=&\frac{1}{\sqrt{V}}\sum_{{\bf p}}\alpha _{b{\bf p}%
}e^{i{\bf p\cdot r}/\hbar },
\end{eqnarray}
where $V$ is the quantization volume and $[\alpha _{i{\bf p}},\alpha _{j{\bf %
p}^{\prime }}^{\dagger }]=\delta _{ij}\delta _{{\bf p},{\bf p}^{\prime }}$
and $[\alpha _{i{\bf p}},\alpha _{j{\bf p}^{\prime }}]=0$. In this basis,
the Hamiltonian becomes 
\end{mathletters}
\begin{mathletters}
\begin{eqnarray}
\hat{H}_{atom} &=&\sum_{{\bf p}}\left\{ \left( \frac{{\bf p}^{2}}{2m}+\frac{%
\hbar \delta }{2}\right) \alpha _{a{\bf p}}^{\dagger }\alpha _{a{\bf p}%
}+\left( \frac{{\bf p}^{2}}{2m}-\frac{\hbar \delta }{2}\right) \alpha _{b%
{\bf p}}^{\dagger }\alpha _{b{\bf p}}+\hbar \Omega _{R}\left( \alpha _{b{\bf %
p}}^{\dagger }\alpha _{a{\bf p}}+\alpha _{a{\bf p}}^{\dagger }\alpha _{b{\bf %
p}}\right) \right\} ,  \label{ha1} \\
\hat{H}_{coll} &=&\frac{1}{2V}\sum_{{\bf p}_{1}+{\bf p}_{2}={\bf p}_{3}+{\bf %
p}_{4}}\left\{ U_{aa}\alpha _{a{\bf p}_{1}}^{\dagger }\alpha _{a{\bf p}%
_{2}}^{\dagger }\alpha _{a{\bf p}_{3}}\alpha _{a{\bf p}_{4}}+U_{bb}\alpha _{b%
{\bf p}_{1}}^{\dagger }\alpha _{b{\bf p}_{2}}^{\dagger }\alpha _{b{\bf p}%
_{3}}\alpha _{b{\bf p}_{4}}+2U_{ab}\alpha _{a{\bf p}_{1}}^{\dagger }\alpha
_{b{\bf p}_{2}}^{\dagger }\alpha _{a{\bf p}_{3}}\alpha _{b{\bf p}%
_{4}}\right\} .  \label{ha2}
\end{eqnarray}

At this point it is advantageous to introduce the grand canonical
Hamiltonian, $\hat{K}=\hat{H}-\mu \hat{N}$ where $\mu $ is the chemical
potential and $\hat{N}=\sum_{{\bf p}}\left( \alpha _{a{\bf p}}^{\dagger
}\alpha _{a{\bf p}}+\alpha _{b{\bf p}}^{\dagger }\alpha _{b{\bf p}}\right) $
is the number operator. The motivation for using $\hat{K}$ is that the
Bogoliubov prescription which will be used later, results in a Hamiltonian
that does not conserve the number of particles. The chemical potential
serves as a Lagrange multiplier which allows one to impose the constraint $%
\left\langle \hat{N}\right\rangle =N$ \cite{fetter}. On the other hand,
since $\mu $ is the energy of an atom in the condensate ground state, $\hat{K%
}$ corresponds to a representation in which all single particle energies are
measured relative to the condensate.

Dressed operators $c_{{\bf p}}$ and $d_{{\bf p}}$, which correspond to the
atomic dressed states $\left| c\right\rangle =\cos \theta \left|
a\right\rangle +\sin \theta \left| b\right\rangle $ and $\left|
d\right\rangle =-\sin \theta \left| a\right\rangle +\cos \theta \left|
b\right\rangle $, may be introduced which diagonalize $\hat{H}_{atom}-\mu 
\hat{N}$. They are related to the operators $\alpha _{b{\bf p}}$ and $\alpha
_{a{\bf p}}$by 
\end{mathletters}
\begin{equation}
\left( 
\begin{array}{c}
c_{{\bf p}} \\ 
d_{{\bf p}}
\end{array}
\right) =\left( 
\begin{array}{cc}
\cos \theta & \sin \theta \\ 
-\sin \theta & \cos \theta
\end{array}
\right) \left( 
\begin{array}{c}
\alpha _{a{\bf p}} \\ 
\alpha _{b{\bf p}}
\end{array}
\right) ;
\end{equation}
where the dressed state angle $\theta $ is given by $\cos \theta =\frac{1}{%
\sqrt{2}}\left( 1+\frac{\delta }{\sqrt{\delta ^{2}+4\Omega _{R}^{2}}}\right)
^{1/2}$ or $\tan 2\theta =\frac{2\Omega _{R}}{\delta }$. Note that $\theta $
is a measure of the relative mixing of the atomic states $\left|
a\right\rangle $ and $\left| b\right\rangle $ in the dressed states so that,
for example, $\theta =\pi /4$ ($\Omega _{R}\gg \left| \delta \right| $)
corresponds to dressed states which are equal superpositions of the states $%
\left| a\right\rangle $ and $\left| b\right\rangle .$ In contrast, for $%
\theta =0$ ($\delta >0$ and $\Omega _{R}\ll \left| \delta \right| $) one has 
$\left| c\right\rangle =\left| a\right\rangle $ and for $\theta =\pi /2$ ( $%
\delta <0$ and $\Omega _{R}\ll \left| \delta \right| $) one has $\left|
c\right\rangle =\left| b\right\rangle $ so that in these limits, the dressed
states may be identified with the atomic states. It is easy to show that $%
[c_{{\bf p}},c_{{\bf p}^{\prime }}^{\dagger }]=[d_{{\bf p}},d_{{\bf p}%
^{\prime }}^{\dagger }]=\delta _{{\bf p},{\bf p}^{\prime }}$ and $[c_{{\bf p}%
},c_{{\bf p}^{\prime }}]=[d_{{\bf p}},d_{{\bf p}^{\prime }}]=[d_{{\bf p}},c_{%
{\bf p}^{\prime }}^{\dagger }]=[d_{{\bf p}},c_{{\bf p}^{\prime }}]=0.$

In the dressed basis $\hat{H}_{atom}-\mu \hat{N}$ has the form 
\begin{equation}
\hat{H}_{atom}-\mu \hat{N}=\sum_{{\bf p}}\left\{ \left( \frac{{\bf p}^{2}}{2m%
}+\frac{\hbar \omega _{cd}}{2}-\mu \right) c_{{\bf p}}^{\dagger }c_{{\bf p}%
}+\left( \frac{{\bf p}^{2}}{2m}-\frac{\hbar \omega _{cd}}{2}-\mu \right) d_{%
{\bf p}}^{\dagger }d_{{\bf p}}\right\} ;
\end{equation}
where $\hbar \omega _{cd}=\hbar \sqrt{\delta ^{2}+4\Omega _{R}^{2}}$ is the
energy difference between the two dressed states. However, in the dressed
basis $\hat{H}_{coll}$ has a significantly more complicated form, 
\begin{eqnarray}
\hat{H}_{coll} &=&\frac{1}{2V}\sum_{{\bf p}_{1}+{\bf p}_{2}={\bf p}_{3}+{\bf %
p}_{4}}\left\{ U_{1}c_{{\bf p}_{1}}^{\dagger }c_{{\bf p}_{2}}^{\dagger }c_{%
{\bf p}_{3}}c_{{\bf p}_{4}}+U_{2}d_{{\bf p}_{1}}^{\dagger }d_{{\bf p}%
_{2}}^{\dagger }d_{{\bf p}_{3}}d_{{\bf p}_{4}}+U_{3}c_{{\bf p}_{1}}^{\dagger
}d_{{\bf p}_{2}}^{\dagger }c_{{\bf p}_{3}}d_{{\bf p}_{4}}\right.  \nonumber
\\
&&\left. +U_{4}\left( c_{{\bf p}_{1}}^{\dagger }c_{{\bf p}_{2}}^{\dagger }d_{%
{\bf p}_{3}}d_{{\bf p}_{4}}+d_{{\bf p}_{1}}^{\dagger }d_{{\bf p}%
_{2}}^{\dagger }c_{{\bf p}_{3}}c_{{\bf p}_{4}}\right) +U_{5}\left( c_{{\bf p}%
_{1}}^{\dagger }c_{{\bf p}_{2}}^{\dagger }c_{{\bf p}_{3}}d_{{\bf p}_{4}}+d_{%
{\bf p}_{1}}^{\dagger }c_{{\bf p}_{2}}^{\dagger }c_{{\bf p}_{3}}c_{{\bf p}%
_{4}}\right) \right.  \nonumber \\
&&\left. +U_{6}\left( c_{{\bf p}_{1}}^{\dagger }d_{{\bf p}_{2}}^{\dagger }d_{%
{\bf p}_{3}}d_{{\bf p}_{4}}+d_{{\bf p}_{1}}^{\dagger }d_{{\bf p}%
_{2}}^{\dagger }d_{{\bf p}_{3}}c_{{\bf p}_{4}}\right) \right\} .
\end{eqnarray}
The collisional couplings $U_{i}$ are given by, 
\begin{mathletters}
\begin{eqnarray}
U_{1} &=&U_{aa}\cos ^{4}\theta +U_{bb}\sin ^{4}\theta +\frac{1}{2}U_{ab}\sin
^{2}2\theta ; \\
U_{2} &=&U_{aa}\sin ^{4}\theta +U_{bb}\cos ^{4}\theta +\frac{1}{2}U_{ab}\sin
^{2}2\theta ; \\
U_{3} &=&(U_{aa}+U_{bb})\sin ^{2}2\theta +2U_{ab}\cos ^{2}2\theta ; \\
U_{4} &=&\frac{1}{4}(U_{aa}+U_{bb}-2U_{ab})\sin ^{2}2\theta ; \\
U_{5} &=&\sin 2\theta (U_{bb}\sin ^{2}\theta -U_{aa}\cos ^{2}\theta
+U_{ab}\cos 2\theta ); \\
U_{6} &=&\sin 2\theta (U_{bb}\cos ^{2}\theta -U_{aa}\sin ^{2}\theta
-U_{ab}\cos 2\theta ).
\end{eqnarray}

Even though the atom-radiation field interaction term has been eliminated,
it would seem that nothing is gained by using the dressed basis because of
the increased complexity of $\hat{H}_{coll}.$ However, one may make a
transformation to an interaction representation with respect to the internal
dressed state energies by defining new slowly varying operators $\bar{c}_{%
{\bf p}}$ and $\bar{d}_{{\bf p}}$ which are related to the normal dressed
operators by, 
\end{mathletters}
\begin{mathletters}
\begin{eqnarray}
c_{{\bf p}} &=&\bar{c}_{{\bf p}}e^{-i\omega _{cd}t/2}; \\
d_{{\bf p}} &=&\bar{d}_{{\bf p}}e^{+i\omega _{cd}t/2}.
\end{eqnarray}
The time evolution of $\bar{c}_{{\bf p}}$ and $\bar{d}_{{\bf p}}$ is
governed by the Hamiltonian, $\hat{K}_{s}$, given by 
\end{mathletters}
\begin{eqnarray}
\hat{K}_{s} &=&\sum_{{\bf p}}\left\{ \left( \frac{{\bf p}^{2}}{2m}-\mu
\right) \bar{c}_{{\bf p}}^{\dagger }\bar{c}_{{\bf p}}+\left( \frac{{\bf p}%
^{2}}{2m}-\mu \right) \bar{d}_{{\bf p}}^{\dagger }\bar{d}_{{\bf p}}\right\} +
\nonumber \\
&&\frac{1}{2V}\sum_{{\bf p}_{1}+{\bf p}_{2}={\bf p}_{3}+{\bf p}_{4}}\left\{
U_{1}\bar{c}_{{\bf p}_{1}}^{\dagger }\bar{c}_{{\bf p}_{2}}^{\dagger }\bar{c}%
_{{\bf p}_{3}}\bar{c}_{{\bf p}_{4}}+U_{2}\bar{d}_{{\bf p}_{1}}^{\dagger }%
\bar{d}_{{\bf p}_{2}}^{\dagger }\bar{d}_{{\bf p}_{3}}\bar{d}_{{\bf p}%
_{4}}+U_{3}\bar{c}_{{\bf p}_{1}}^{\dagger }\bar{d}_{{\bf p}_{2}}^{\dagger }%
\bar{c}_{{\bf p}_{3}}\bar{d}_{{\bf p}_{4}}\right.  \nonumber \\
&&\left. +U_{4}\left( e^{2i\omega _{cd}t}\bar{c}_{{\bf p}_{1}}^{\dagger }%
\bar{c}_{{\bf p}_{2}}^{\dagger }\bar{d}_{{\bf p}_{3}}\bar{d}_{{\bf p}%
_{4}}+e^{-2i\omega _{cd}t}\bar{d}_{{\bf p}_{1}}^{\dagger }\bar{d}_{{\bf p}%
_{2}}^{\dagger }\bar{c}_{{\bf p}_{3}}\bar{c}_{{\bf p}_{4}}\right)
+U_{5}\left( e^{i\omega _{cd}t}\bar{c}_{{\bf p}_{1}}^{\dagger }\bar{c}_{{\bf %
p}_{2}}^{\dagger }\bar{c}_{{\bf p}_{3}}\bar{d}_{{\bf p}_{4}}+e^{-i\omega
_{cd}t}\bar{d}_{{\bf p}_{1}}^{\dagger }\bar{c}_{{\bf p}_{2}}^{\dagger }\bar{c%
}_{{\bf p}_{3}}\bar{c}_{{\bf p}_{4}}\right) \right.  \nonumber \\
&&\left. +U_{6}\left( e^{i\omega _{cd}t}\bar{c}_{{\bf p}_{1}}^{\dagger }\bar{%
d}_{{\bf p}_{2}}^{\dagger }\bar{d}_{{\bf p}_{3}}\bar{d}_{{\bf p}%
_{4}}+e^{-i\omega _{cd}t}\bar{d}_{{\bf p}_{1}}^{\dagger }\bar{d}_{{\bf p}%
_{2}}^{\dagger }\bar{d}_{{\bf p}_{3}}\bar{c}_{{\bf p}_{4}}\right) \right\} .
\label{Ks}
\end{eqnarray}
The terms proportional to $U_{4}$, $U_{5}$, and $U_{6}$ oscillate with a
frequency that is a multiple of $\omega _{cd}$. The time scales over which
the operators $\bar{c}_{{\bf p}}$ and $\bar{d}_{{\bf p}}$ evolve are
governed by the single particle kinetic energy, $\frac{{\bf p}^{2}}{2m}$,
and the mean-field interaction energies, $n_{o}U_{i}$, which are on the
order of $n_{o}U_{aa}$, $n_{o}U_{bb}$, and $n_{o}U_{ab}$ where $n_{o}$ is
the density of atoms in the condensate. Consequently, when the dressed state
energy splitting is sufficiently large so that 
\begin{equation}
\hbar \omega _{cd}\gg \frac{{\bf p}^{2}}{2m}%
,n_{o}U_{aa},n_{o}U_{bb},n_{o}U_{ab},  \label{split}
\end{equation}
the oscillatory terms in $\hat{K}_{s}$ will undergo many oscillations before 
$\bar{c}_{{\bf p}}$ and $\bar{d}_{{\bf p}}$ will have changed significantly.
One may therefore consider the Heisenberg equation of motion for $\bar{c}_{%
{\bf p}}$ which is averaged over a time interval $T=2\pi /\omega _{cd}$, 
\begin{equation}
\frac{1}{T}\int_{t}^{t+T}\frac{d\bar{c}_{{\bf p}}}{dt}dt=\frac{1}{T}%
\int_{t}^{t+T}\frac{1}{i\hbar }[\bar{c}_{{\bf p}},\hat{K}_{s}]dt.
\label{coarse}
\end{equation}
This allows one to define a coarse-grained derivative, 
\begin{equation}
\frac{\delta \bar{c}_{{\bf p}}}{\delta T}=\frac{1}{T}\int_{t}^{t+T}\frac{d%
\bar{c}_{{\bf p}}}{dt}dt=\frac{\bar{c}_{{\bf p}}(t+T)-\bar{c}_{{\bf p}}(t)}{T%
},
\end{equation}
When condition (\ref{split}) is satisfied, the operators in $\hat{K}_{s}$
may be treated as constant over the period $T$ so that 
\begin{equation}
\frac{\delta \bar{c}_{{\bf p}}}{\delta T}=\frac{1}{i\hbar }[\bar{c}_{{\bf p}%
},\hat{K}_{sr}]
\end{equation}
where $\hat{K}_{sr}$ is the ''resonant'' Hamiltonian, 
\begin{eqnarray}
\hat{K}_{sr} &=&\sum_{{\bf p}}\left\{ \left( \frac{{\bf p}^{2}}{2m}-\mu
\right) \bar{c}_{{\bf p}}^{\dagger }\bar{c}_{{\bf p}}+\left( \frac{{\bf p}%
^{2}}{2m}-\mu \right) \bar{d}_{{\bf p}}^{\dagger }\bar{d}_{{\bf p}}\right\} +
\nonumber \\
&&\frac{1}{2V}\sum_{{\bf p}_{1}+{\bf p}_{2}={\bf p}_{3}+{\bf p}_{4}}\left\{
U_{1}\bar{c}_{{\bf p}_{1}}^{\dagger }\bar{c}_{{\bf p}_{2}}^{\dagger }\bar{c}%
_{{\bf p}_{3}}\bar{c}_{{\bf p}_{4}}+U_{2}\bar{d}_{{\bf p}_{1}}^{\dagger }%
\bar{d}_{{\bf p}_{2}}^{\dagger }\bar{d}_{{\bf p}_{3}}\bar{d}_{{\bf p}%
_{4}}+U_{3}\bar{c}_{{\bf p}_{1}}^{\dagger }\bar{d}_{{\bf p}_{2}}^{\dagger }%
\bar{c}_{{\bf p}_{3}}\bar{d}_{{\bf p}_{4}}\right\} .  \label{resonant}
\end{eqnarray}
Since $\bar{c}_{{\bf p}}$ changes very little in a time $T$ one may identify
the coarse grained derivative with the actual derivative, 
\begin{equation}
\frac{\delta \bar{c}_{{\bf p}}}{\delta T}\approx \frac{d\bar{c}_{{\bf p}}}{dt%
}.
\end{equation}
The same results hold for $\bar{d}_{{\bf p}}.$ As a result, one may consider
the time evolution of $\bar{c}_{{\bf p}}$ and $\bar{d}_{{\bf p}}$ to be
governed by $\hat{K}_{sr}$.

Neglecting the oscillatory terms in $\hat{K}_{s}$ is analogous to the
rotating wave approximation in quantum optics \cite{rwa}\cite{allen} and has
a simple interpretation in terms of energy conservation. As an example,
consider the term $\bar{c}_{{\bf p}_{1}}^{\dagger }\bar{c}_{{\bf p}%
_{2}}^{\dagger }\bar{d}_{{\bf p}_{3}}\bar{d}_{{\bf p}_{4}}$in $\hat{K}_{s}$.
This term corresponds to the destruction of two atoms in the dressed state $%
\left| d\right\rangle $ and the creation of two atoms in state $\left|
c\right\rangle $ which requires an amount of energy equal to $2\hbar \omega
_{cd}.$ However when condition (\ref{split}) is satisfied, the kinetic
energy and mean-field energy of the atoms is insufficient to overcome the
energy difference of $2\hbar \omega _{cd}.$ Consequently, the process
described by $\bar{c}_{{\bf p}_{1}}^{\dagger }\bar{c}_{{\bf p}_{2}}^{\dagger
}\bar{d}_{{\bf p}_{3}}\bar{d}_{{\bf p}_{4}}$ can only occur over time scales
consistent with the uncertainty relation $\Delta t\Delta E\approx \hbar ,$
which in this case is $\Delta t\approx 1/2\omega _{cd}.$ However, by coarse
graining the time evolution of the operators $\bar{c}_{{\bf p}}$ and $\bar{d}%
_{{\bf p}}$, one ignores processes that occur on time scales less than $%
T>\Delta t.$

The slowly varying operators were useful for showing that several of the
terms $\hat{H}_{coll}$ could be neglected. However, for the remainder of the
paper, all calculations will be carried out using the operators $c_{{\bf p}}$
and $d_{{\bf p}}$ in the normal Heisenberg representation with the
''resonant'' Hamiltonian $\hat{K}_{R}$ given by, 
\begin{eqnarray}
\hat{K}_{R} &=&\sum_{{\bf p}}\left\{ \left( \frac{{\bf p}^{2}}{2m}+\frac{%
\hbar \omega _{cd}}{2}-\mu \right) c_{{\bf p}}^{\dagger }c_{{\bf p}}+\left( 
\frac{{\bf p}^{2}}{2m}-\frac{\hbar \omega _{cd}}{2}-\mu \right) d_{{\bf p}%
}^{\dagger }d_{{\bf p}}\right\} +  \nonumber \\
&&\frac{1}{2V}\sum_{{\bf p}_{1}+{\bf p}_{2}={\bf p}_{3}+{\bf p}_{4}}\left\{
U_{1}c_{{\bf p}_{1}}^{\dagger }c_{{\bf p}_{2}}^{\dagger }c_{{\bf p}_{3}}c_{%
{\bf p}_{4}}+U_{2}d_{{\bf p}_{1}}^{\dagger }d_{{\bf p}_{2}}^{\dagger }d_{%
{\bf p}_{3}}d_{{\bf p}_{4}}+U_{3}c_{{\bf p}_{1}}^{\dagger }d_{{\bf p}%
_{2}}^{\dagger }c_{{\bf p}_{3}}d_{{\bf p}_{4}}\right\} .  \label{Kr}
\end{eqnarray}
Notice that Eq.(\ref{Kr}) resembles Eq. (\ref{ha1}-\ref{ha2}) except that
the external field coupling term is absent. Equation (\ref{Kr}) is the
central result of this paper. The coupling constants $U_{1},$ $U_{2},$ and $%
U_{3}$ in Eq. (\ref{Kr}) are explicit functions of the dressed state angle.
Consequently, the collective properties of the condensate which depend on
the two-body interactions will be a function of $\theta $. In the following
section, the condensate ground state and elementary excitations above the
ground states are calculated using Eq. (\ref{Kr}).

\section{Condensate and Excited States}

The c-number equations for the wave function of the condensate ground state
may be derived by dividing the field operators $\hat{\Psi}_{c}({\bf r})$ and 
$\hat{\Psi}_{d}({\bf r})$, in the dressed basis, into a part with a non-zero
expectation value and a fluctuating part, $\delta \hat{\Psi}_{i}({\bf r})$,
that has a vanishing expectation value with respect to the ground state, 
\begin{mathletters}
\begin{eqnarray}
\hat{\Psi}_{c}({\bf r}) &=&\frac{c_{0}}{\sqrt{V}}+\delta \hat{\Psi}_{c}({\bf %
r}); \\
\hat{\Psi}_{d}({\bf r}) &=&\frac{d_{0}}{\sqrt{V}}+\delta \hat{\Psi}_{d}({\bf %
r}).
\end{eqnarray}
The wave functions for the condensate ground state are $\phi _{c}\equiv
\left\langle \hat{\Psi}_{c}\right\rangle =\frac{\left\langle
c_{0}\right\rangle }{\sqrt{V}}$ and $\phi _{d}\equiv \left\langle \hat{\Psi}%
_{d}\right\rangle =\frac{\left\langle d_{0}\right\rangle }{\sqrt{V}}.$ A
pair of coupled equations for $\phi _{c}$, $\phi _{d},$ and $\mu $ may be
derived from the expectation value of the Heisenberg equations of motion and
using the fact that $\dot{\phi}_{c}=\dot{\phi}_{d}=0$ with respect to $\hat{K%
}_{R}$ \cite{fetter}, 
\end{mathletters}
\begin{mathletters}
\begin{eqnarray}
\mu \phi _{c} &=&(\frac{\hbar }{2}\omega _{cd}+U_{1}|\phi _{c}|^{2}+\frac{1}{%
2}U_{3}|\phi _{d}|^{2})\phi _{c};  \label{ground1} \\
\mu \phi _{d} &=&(-\frac{\hbar }{2}\omega _{cd}+U_{2}|\phi _{d}|^{2}+\frac{1%
}{2}U_{3}|\phi _{c}|^{2})\phi _{d};  \label{ground2}
\end{eqnarray}
with the constraint $n_{o}\equiv \frac{N_{o}}{V}=|\phi _{c}|^{2}+|\phi
_{d}|^{2}$. Equations (\ref{ground1}-\ref{ground2}) may also be derived by
requiring that the energy of the ground state be an extremum with respect to
variations in $\phi _{c}$ and $\phi _{d}$.

Due to the non-zero mean-field interactions, there are three possible
solutions to Eqs. (\ref{ground1}- \ref{ground2}): 
\end{mathletters}
\begin{eqnarray*}
\text{(i) }|\phi _{c}|^{2} &=&n_{o}\text{, }\phi _{d}=0\text{, and }\mu _{c}=%
\frac{\hbar }{2}\omega _{cd}+U_{1}n_{o}; \\
\text{(ii) }|\phi _{d}|^{2} &=&n_{o}\text{, }\phi _{c}=0\text{, and }\mu
_{d}=-\frac{\hbar }{2}\omega _{cd}+U_{2}n_{o}; \\
\text{(iii) }|\phi _{c}|^{2} &=&\frac{(U_{2}-\frac{1}{2}U_{3})n_{o}-\hbar
\omega _{cd}}{U_{1}+U_{2}-U_{3}}\text{, }|\phi _{d}|^{2}=\frac{(U_{1}-\frac{1%
}{2}U_{3})n_{o}+\hbar \omega _{cd}}{U_{1}+U_{2}-U_{3}}\text{, } \\
\text{and }\mu &=&\frac{1}{2}\frac{\left( 2U_{1}U_{2}-\frac{1}{2}%
U_{3}^{2}\right) n_{o}+(U_{2}-U_{1})\hbar \omega _{cd}}{U_{1}+U_{2}-U_{3}}.
\end{eqnarray*}
The derivation of solution (iii) requires $\phi _{d}\neq 0$ and $\phi
_{c}\neq 0$, but it is easy to show that in the limit that $|\phi
_{d}|^{2}\rightarrow 0$ or $|\phi _{c}|^{2}\rightarrow 0$ one recovers
solutions (i) or (ii), respectively. When (\ref{split}) is satisfied, (iii)
is an unphysical solution since either $|\phi _{c}|^{2}<0$ {\it or} $|\phi
_{d}|^{2}<0$ and consequently, case (iii) may be ignored. The solutions (i)
and (ii) correspond to atoms of the condensate being in one of the dressed
states, $\left| c\right\rangle $ or $\left| d\right\rangle $. Case (ii)
corresponds to the thermodynamic ground state since it has the lowest ground
state energy. However, in what follows the excitation spectrum for both
cases is calculated.

The Bogoliubov approximation \cite{bogo} may applied to $\hat{K}_{R}$ for
the two condensate solutions (i) and (ii) to obtain a Hamiltonian which is
quadratic in the operators for the excited states (those states with ${\bf p}%
\neq 0$). For case (i), one obtains the linearized Hamiltonian, 
\begin{eqnarray}
\hat{K}_{Rc} &=&K_{oc}+\sum_{{\bf p}\neq 0}\left\{ \left( \frac{{\bf p}^{2}}{%
2m}+U_{1}n_{o}\right) c_{{\bf p}}^{\dagger }c_{{\bf p}}+\frac{1}{2}%
U_{1}n_{o}\left( c_{{\bf p}}^{\dagger }c_{-{\bf p}}^{\dagger }+c_{{\bf p}%
}c_{-{\bf p}}\right) \right\}  \nonumber \\
&&+\sum_{{\bf p}\neq 0}\left( \frac{{\bf p}^{2}}{2m}-\hbar \omega
_{cd}-(U_{1}-\frac{1}{2}U_{3})n_{o}\right) d_{{\bf p}}^{\dagger }d_{{\bf p}};
\label{Krc}
\end{eqnarray}
by taking $c_{0}^{\dagger }\approx c_{0}=\sqrt{N_{o}}$ and $d_{0}^{\dagger
}\approx d_{0}=0$ and neglecting terms which are cubic and quartic in
operators with ${\bf p}\neq 0.$ Here $K_{oc}=E_{oc}-\mu _{c}N_{o}$ and $%
E_{oc}=\frac{\hbar \omega _{cd}}{2}N_{o}+\frac{1}{2V}U_{1}N_{o}^{2}$ is the
total energy of the ground state.

One can see directly from Eq. (\ref{Krc}) that the excitations which are in
the atomic state that is orthogonal to the ground state, i.e. $\left|
d\right\rangle $, are single particle excitations with an energy spectrum of 
\begin{equation}
\hbar \omega _{d}({\bf p})=\frac{{\bf p}^{2}}{2m}-\hbar \omega _{cd}-(U_{1}-%
\frac{1}{2}U_{3})n_{o}.
\end{equation}
This spectrum corresponds to the free atoms in state $\left| d\right\rangle $
with a mean field shift of $\frac{1}{2}U_{3}n_{o}$ due to the interactions
with the condensate. The term $-\left( \hbar \omega _{cd}+U_{1}n_{o}\right) $
in $\omega _{d}({\bf p})$ results from the fact that the single particle
energies are measured relative to $\mu _{c}$. The reason $\omega _{d}({\bf p}%
)<0$ is due to the fact that ground state (ii) is the global minimum in the
condensate energy.

For excitations in state $\left| c\right\rangle ,$ one can carry out a
canonical transformation by defining quasiparticle operators, $C_{{\bf p}}$,
such that 
\begin{equation}
\sum_{{\bf p}\neq 0}\left\{ \left( \frac{{\bf p}^{2}}{2m}+U_{1}n_{o}\right)
c_{{\bf p}}^{\dagger }c_{{\bf p}}+\frac{1}{2}U_{1}n_{o}\left( c_{{\bf p}%
}^{\dagger }c_{-{\bf p}}^{\dagger }+c_{{\bf p}}c_{-{\bf p}}\right) \right\}
=\sum_{{\bf p}\neq 0}\hbar \omega _{c}({\bf p})C_{{\bf p}}^{\dagger }C_{{\bf %
p}}+E_{vac,c};
\end{equation}
where $E_{vac,c}$ is the vacuum energy for the quasiparticle vacuum. The $C_{%
{\bf p}}$ obey the bosonic commutation relations $[C_{{\bf p}},C_{{\bf p}%
^{\prime }}^{\dagger }]=\delta _{{\bf p},{\bf p}^{\prime }}$ and $[C_{{\bf p}%
},C_{{\bf p}^{\prime }}]=0$ and can be expressed in terms $c_{{\bf p}}$ and $%
c_{{\bf p}}^{\dagger }$ as 
\begin{equation}
C_{{\bf p}}=\cosh \varphi _{{\bf p}}c_{{\bf p}}+\sinh \varphi _{{\bf p}}c_{-%
{\bf p}}^{\dagger };  \label{quasi}
\end{equation}
The solutions for the quasiparticle energies and $\varphi _{{\bf p}}$ are
easily found to be \cite{fetter}, 
\begin{mathletters}
\begin{eqnarray}
\hbar \omega _{c}({\bf p}) &=&\sqrt{\frac{{\bf p}^{2}}{2m}\left( \frac{{\bf p%
}^{2}}{2m}+2U_{1}n_{o}\right) }; \\
\tanh 2\varphi _{{\bf p}} &=&\frac{U_{1}n_{o}}{\frac{{\bf p}^{2}}{2m}%
+U_{1}n_{o}}.
\end{eqnarray}
The long-wavelength quasiparticle excitations for which $\frac{{\bf p}^{2}}{%
2m}\ll 2U_{1}n_{o}$ correspond to phonons with the dispersion relation $%
\hbar \omega _{c}({\bf p})=u_{c}p$ and a speed of sound $u_{c}$ given by 
\end{mathletters}
\begin{equation}
u_{c}=\sqrt{\frac{U_{1}n_{o}}{m}}=\sqrt{\frac{n_{o}}{m}\left( U_{aa}\cos
^{4}\theta +U_{bb}\sin ^{4}\theta +\frac{1}{2}U_{ab}\sin ^{2}2\theta \right) 
};  \label{sound2}
\end{equation}
The speed of sound given by Eq.(\ref{sound2}) is the same as that calculated
from the pressure, $P$, of the condensate using $P=-\frac{\partial E_{oc}}{%
\partial V}$ along with $u_{c}^{2}=\frac{1}{m}\frac{\partial P}{\partial
n_{o}}$. It is easy to show that the excitations in state $\left|
c\right\rangle $ give rise to density perturbations, $\delta \hat{n}({\bf r}%
,t).$ The number density operator is 
\begin{equation}
\hat{n}({\bf r})=\hat{\Psi}_{c}^{\dagger }({\bf r})\hat{\Psi}_{c}({\bf r})+%
\hat{\Psi}_{d}^{\dagger }({\bf r})\hat{\Psi}_{d}({\bf r})=\frac{1}{V}\sum_{%
{\bf p},{\bf p}^{\prime }}\left( c_{{\bf p}^{\prime }}^{\dagger }c_{{\bf p}%
}+d_{{\bf p}^{\prime }}^{\dagger }d_{{\bf p}}\right) e^{i({\bf p}-{\bf p}%
^{\prime }){\bf \cdot r}/\hbar }
\end{equation}
which may be linearized around the condensate ground state to give $\hat{n}(%
{\bf r})\approx n_{o}+\delta \hat{n}({\bf r},t)$. The density perturbations
can be expressed in terms of the quasiparticles as 
\begin{equation}
\delta \hat{n}({\bf r},t)=\left( \frac{n_{o}}{V}\right) ^{1/2}\sum_{{\bf p}%
\neq 0}\left( \cosh \varphi _{{\bf p}}-\sinh \varphi _{{\bf p}}\right)
\left( C_{{\bf p}}^{\dagger }(t)e^{-i{\bf p\cdot r}/\hbar }+C_{{\bf p}%
}(t)e^{i{\bf p\cdot r}/\hbar }\right) ,
\end{equation}
which has the form for the phonon states, $p\ll mu_{c},$ of 
\begin{equation}
\delta \hat{n}({\bf r},t)=\sum_{p\ll mu_{c}}\left( \frac{n_{o}p}{2Vmu_{c}}%
\right) ^{1/2}\left( C_{{\bf p}}^{\dagger }(0)e^{-i({\bf p\cdot r-}%
u_{c}pt)/\hbar }+C_{{\bf p}}(0)e^{i({\bf p\cdot r-}u_{c}pt)/\hbar }\right) .
\label{phonon}
\end{equation}
Equation (\ref{phonon}) is the operator for the density perturbations which
obeys Eq.(\ref{wave}) for a homogenous fluid \cite{fetter}. Consequently,
the excitations in state $\left| d\right\rangle $ do not contribute (at
least to order $\sqrt{N_{o}}$) to sound propagation.

The calculation for case (ii) proceeds in an identical manner and, as such,
we quote only the main results. The linearized Hamiltonian is, 
\begin{eqnarray}
\hat{K}_{Rd} &=&K_{od}+\sum_{{\bf p}\neq 0}\left\{ \left( \frac{{\bf p}^{2}}{%
2m}+U_{2}n_{o}\right) d_{{\bf p}}^{\dagger }d_{{\bf p}}+\frac{1}{2}%
U_{2}n_{o}\left( d_{{\bf p}}^{\dagger }d_{-{\bf p}}^{\dagger }+d_{{\bf p}%
}d_{-{\bf p}}\right) \right\}  \nonumber \\
&&+\sum_{{\bf p}\neq 0}\left( \frac{{\bf p}^{2}}{2m}+\hbar \omega
_{cd}-(U_{2}-\frac{1}{2}U_{3})n_{o}\right) c_{{\bf p}}^{\dagger }c_{{\bf p}};
\label{Krd}
\end{eqnarray}
with $K_{od}=E_{od}-\mu N_{o}$ and $E_{od}=-\frac{\hbar \omega _{cd}}{2}%
N_{o}+\frac{1}{2V}U_{2}N_{o}^{2}$ is the ground state energy. The
excitations in state $\left| c\right\rangle $ are single particle
excitations with an energy spectrum of $\hbar \omega _{c}({\bf p})=\frac{%
{\bf p}^{2}}{2m}+\hbar \omega _{cd}-(U_{2}-\frac{1}{2}U_{3})n_{o}$. The
quasiparticle operators $D_{{\bf p}}=\cosh \tilde{\varphi}_{{\bf p}}d_{{\bf p%
}}+\sinh \tilde{\varphi}_{{\bf p}}d_{-{\bf p}}^{\dagger }$ diagonalize the
terms in $\hat{K}_{Rd}$ involving $d_{{\bf p}}$ and $d_{{\bf p}}^{\dagger }$%
, 
\begin{equation}
\sum_{{\bf p}\neq 0}\left\{ \left( \frac{{\bf p}^{2}}{2m}+U_{2}n_{o}\right)
d_{{\bf p}}^{\dagger }d_{{\bf p}}+\frac{1}{2}U_{2}n_{o}\left( d_{{\bf p}%
}^{\dagger }d_{-{\bf p}}^{\dagger }+d_{{\bf p}}d_{-{\bf p}}\right) \right\}
=\sum_{{\bf p}\neq 0}\hbar \omega _{d}({\bf p})D_{{\bf p}}^{\dagger }D_{{\bf %
p}}+E_{vac,d}
\end{equation}
with quasiparticle energies 
\begin{equation}
\hbar \omega _{d}({\bf p})=\sqrt{\frac{{\bf p}^{2}}{2m}\left( \frac{{\bf p}%
^{2}}{2m}+2U_{2}n_{o}\right) }.
\end{equation}
The long wavelength phonon excitations have a sound velocity $u_{d}$ given
by 
\begin{equation}
u_{d}=\sqrt{\frac{U_{2}n_{o}}{m}}=\sqrt{\frac{n_{o}}{m}\left( U_{aa}\sin
^{4}\theta +U_{bb}\cos ^{4}\theta +\frac{1}{2}U_{ab}\sin ^{2}2\theta \right) 
}.  \label{sound3}
\end{equation}
Again, the sound speed given by Eq. (\ref{sound3}) is identical to that
calculated using $P=-\frac{\partial E_{od}}{\partial V}$ and $u_{d}^{2}=%
\frac{1}{m}\frac{\partial P}{\partial n_{o}}.$ The linearized density
perturbations are now given by, 
\begin{eqnarray}
\delta \hat{n}({\bf r},t) &=&\left( \frac{n_{o}}{V}\right) ^{1/2}\sum_{{\bf p%
}\neq 0}\left( \cosh \tilde{\varphi}_{{\bf p}}-\sinh \tilde{\varphi}_{{\bf p}%
}\right) \left( D_{{\bf p}}^{\dagger }(t)e^{-i{\bf p\cdot r}/\hbar }+D_{{\bf %
p}}(t)e^{i{\bf p\cdot r}/\hbar }\right) ;  \nonumber \\
&=&\sum_{p\ll mu_{d}}\left( \frac{n_{o}p}{2Vmu_{d}}\right) ^{1/2}\left( D_{%
{\bf p}}^{\dagger }(0)e^{-i({\bf p\cdot r-}u_{d}pt)/\hbar }+D_{{\bf p}%
}(0)e^{i({\bf p\cdot r-}u_{d}pt)/\hbar }\right) ;  \label{phonon2}
\end{eqnarray}

Equations (\ref{phonon}) and (\ref{phonon2}) for $\delta \hat{n}({\bf r},t)$
and Eqs. (\ref{sound2}) and (\ref{sound3}) for the speed of sound are the
main results of this section. The expressions for $u_{c}$ and $u_{d}$ given
by Eqs. (\ref{sound2}) and (\ref{sound3}) indicate that the speed at which a
density perturbation propagates in the condensate, depends on the particular
dressed state that the condensate atoms are in {\it and} the value of the
dressed state angle. In the limit that $U_{aa}=U_{bb}=U_{ab}=\frac{4\pi
\hbar ^{2}a}{m},$ one obtains $u_{c}=u_{d}=u$ where $u$ is given by Eq. (\ref
{sound}). Consequently, the inequality of the scattering lengths $a_{ij}$ is
crucial for this effect to be observed. Figures 1(a-c) show plots of $u_{c}$
and $u_{d}$ for various ratios of $U_{bb}/U_{aa}$ and $U_{ab}/U_{aa}$.

The atoms in the condensate can be prepared in either states (i) or (ii) by
adiabatically turning on the external field. Suppose that at $t=-\infty ,$ $%
\Omega _{R}=0$ and $\delta \neq 0$ so that the atoms are in either $\left|
a\right\rangle $ or $\left| b\right\rangle .$ If the atoms initially in
state $\left| b\right\rangle =\left| d\right\rangle $ ($\delta >0$) , then
the atoms will remain in $\left| d\right\rangle $ provided $\theta $ varies
sufficiently slowly. For free atoms, the atoms will adiabatically remain in
the initial dressed state provided $|\dot{\theta}|\omega _{cd}^{-1}\ll 1.$
However, for interacting atoms in a condensate, $\hbar |\dot{\theta}|$
should be mush less than all of the energies in the Heisenberg equations of
motion which govern the time evolution of $c_{{\bf p}}$ and $d_{{\bf p}}.$
Consequently, one must satisfy the more stringent requirement $\hbar \omega
_{cd}\gg \frac{{\bf p}^{2}}{2m},n_{o}U_{aa},n_{o}U_{bb},n_{o}U_{ab}\gg \hbar
|\dot{\theta}|.$

\section{Discussion}

Up to this point, the oscillatory terms in Eq. (\ref{Ks}) have been
neglected under the assumption that their effect is small compared to the
terms in $\hat{K}_{sr}$. However, it is important to obtain an estimate of
the leading order correction due to these terms. The lowest order effect of
the oscillatory terms is an energy shift in the chemical potential and
quasiparticle spectrum which is analogous to the Bloch-Siegert frequency
shift in the atomic resonance of a two-level atom due to the
counter-rotating terms in the atom-field coupling \cite{allen}. The shift in
the chemical potential (i.e. $\mu \rightarrow \mu +\delta \mu $ where $%
\delta \mu $ is the shift) for cases (i) and (ii) may be calculated by
including the two-body interactions proportional to $U_{i}$ for $i=4,5,6$ in
Eqs. (\ref{ground1}-\ref{ground2}). Doing so, one finds that for case (i)
the shift is $\delta \mu _{c}=\frac{3\left( U_{5}n_{o}\right) ^{2}}{4\hbar
\omega _{cd}}$ and for case (ii) the shift is $\delta \mu _{d}=\frac{%
-3\left( U_{6}n_{o}\right) ^{2}}{4\hbar \omega _{cd}}.$

To obtain the shift in the quasiparticle spectrum, one may write Eq. (\ref
{Ks}) as $\hat{K}_{s}=\hat{K}_{sr}+\hat{H}_{cs,nr}(t)$ where $\hat{K}_{sr}$
is given by Eq. (\ref{resonant}) and $\hat{H}_{cs,nr}(t)$ consists of the
terms which oscillate at $\omega _{cd}.$ To calculate the energy shift, one
first linearizes $\hat{H}_{cs,nr}(t)$ around one of the ground states given
by cases (i) or (ii) so that $\hat{H}_{cs,nr}(t)$ is quadratic in $\bar{c}_{%
{\bf p}}$ and $\bar{d}_{{\bf p}}$ for ${\bf p}\neq 0$. Since $e^{i\omega
_{cd}t}$ varies rapidly compared to $\bar{c}_{{\bf p}}$ and $\bar{d}_{{\bf p}%
},$ the coupling between the operators $\bar{c}_{{\bf p}}$ and $\bar{d}_{%
{\bf p}}$ in $\hat{H}_{cs,nr}(t)$ may be adiabatically eliminated by
integrating the Heisenberg equations of motion for the quasiparticles and
substituting the solution back into $\hat{H}_{cs,nr}(t).$ One then finds for
case (i), that Eq. (\ref{Krc}) now has the form $\hat{K}%
_{Rc}=K_{oc}+E_{vac}+\hbar \sum_{{\bf p}\neq 0}\left[ \left( \omega _{c}(%
{\bf p})+\delta \omega _{c}({\bf p})\right) C_{{\bf p}}^{\dagger }C_{{\bf p}%
}+\left( \omega _{d}({\bf p})+\delta \omega _{d}({\bf p})\right) d_{{\bf p}%
}^{\dagger }d_{{\bf p}}\right] $ where the lowest order energy shifts of the
excited states, including the shift in the chemical potential, are given by $%
\hbar \delta \omega _{c}({\bf p})=\frac{\left( U_{5}n_{o}\right) ^{2}}{%
4\hbar \omega _{cd}}\left( \frac{7\frac{{\bf p}^{2}}{2m}-U_{1}n_{o}}{\hbar
\omega _{c}({\bf p})}\right) $ and $\hbar \delta \omega _{d}({\bf p})=\frac{%
\left( U_{4}n_{o}\right) ^{2}-3\left( U_{5}n_{o}\right) ^{2}/2}{2\hbar
\omega _{cd}}$. Note that the factor $\frac{7\frac{{\bf p}^{2}}{2m}%
-U_{1}n_{o}}{\hbar \omega _{c}({\bf p})}$ in $\hbar \delta \omega _{c}({\bf p%
})$ comes from expressing the $\bar{c}_{{\bf p}}$ operators in the
linearized $\hat{H}_{cs,nr}(t)$ in terms of $C_{{\bf p}}$ using Eq. (\ref
{quasi}). For the energy shift to be negligible for the phonon states (i.e. $%
\omega _{c}({\bf p})\gg \delta \omega _{c}({\bf p})$), one must satisfy the
condition $\frac{{\bf p}^{2}}{2m}\gg \frac{\left( U_{5}n_{o}\right) ^{2}}{%
8\hbar \omega _{cd}}.$ In a similar manner one may calculate the energy
shift in the excited states corresponding to ground state (ii). In this case
one finds that the energy shifts in Eq. (\ref{Krd}) are $\hbar \delta \omega
_{c}({\bf p})=-\frac{\left( U_{4}n_{o}\right) ^{2}-3\left( U_{6}n_{o}\right)
^{2}/2}{2\hbar \omega _{cd}}$ and $\hbar \delta \omega _{d}({\bf p})=-\frac{%
\left( U_{6}n_{o}\right) ^{2}}{4\hbar \omega _{cd}}\left( \frac{7\frac{{\bf p%
}^{2}}{2m}-U_{2}n_{o}}{\hbar \omega _{d}({\bf p})}\right) .$ Again, one
finds that for case (ii), the energy shift for the phonon excitations is
negligible provided $\frac{{\bf p}^{2}}{2m}\gg \frac{\left(
U_{6}n_{o}\right) ^{2}}{8\hbar \omega _{cd}}.$ One can see that the energy
shifts are smaller than the unperturbed energies by factors which goes like $%
\frac{U_{i}n_{o}}{\hbar \omega _{cd}}\ll 1$ and that even for the low energy
phonon states, the energy shift is negligible.

The validity of the results in the beginning of this section as well as in
sections II and III all rely on condition (\ref{split}) being satisfied.
Since we are primarily interested in the low energy phonon states for which $%
\frac{{\bf p}^{2}}{2m}\ll n_{o}U_{aa},n_{o}U_{bb},n_{o}U_{ab}$, condition (%
\ref{split}) reduces to 
\begin{equation}
\hbar \omega _{cd}\gg n_{o}U_{aa},n_{o}U_{bb},n_{o}U_{ab}.  \label{splitmean}
\end{equation}
To satisfy this condition as the external field turns on, it is necessary to
have $\hbar \left| \delta \right| \gg n_{o}U_{aa},n_{o}U_{bb},n_{o}U_{ab}.$
At later times, however, once $\Omega _{R}$ is established, the detuning can
be varied adiabatically to adjust the dressed state angle. For these times,
to satisfy (\ref{splitmean}) it is sufficient that 
\begin{equation}
\hbar 2\Omega _{R}\gg n_{o}U_{aa},n_{o}U_{bb},n_{o}U_{ab}.  \label{split2}
\end{equation}
To estimate the mean-field interactions, one can take $a_{ij}\sim a$ and use
the values of $a=2.75nm$ for $^{23}Na$ and $a=5.77nm$ for $^{87}Rb$ and a
condensate density of $n_{o}\sim 10^{14}cm^{-3}$ \cite{stringari}. The
mean-field energy, $\frac{4\pi \hbar ^{2}an_{o}}{m}$, divided by $\hbar $
for $^{23}Na$ is then $9.5\times 10^{3}s^{-1}$ and for $^{87}Rb$ it is $%
5.3\times 10^{3}s^{-1}$. Notice that this is comparable to the two-photon
Rabi frequencies currently used in experiments with $^{87}Rb$ which are
typically $\sim 2\pi \times 600s^{-1}$ \cite{two-photon}.

In order to obtain some feeling of the necessary field strengths needed to
satisfy (\ref{split}), one can, for the sake of definiteness, consider
single photon rf transitions between Zeeman states in a spinor condensate so
that $\hbar \Omega _{R}\sim \mu _{B}B$ where $\mu _{B}$ is the Bohr magneton
and $B$ the magnitude of the rf magnetic field. Thus a magnetic field of $%
B\gg 10^{-3}G$ would be required to satisfy (\ref{split2}). One could also
hold $\delta $ constant and vary the rf field strength $B.$ The splitting
between Zeeman states in the presence of a static uniform magnetic field $%
B_{o}$ will be on the order of $\omega _{o}\sim \mu _{B}B_{o}/\hbar \approx
9\times 10^{10}s^{-1}$ for $B_{o}=10^{4}G.$ One could therefore achieve
detunings such that $\delta \ll \omega _{o}$ so that the rotating wave
approximation remains valid but still have $\hbar \left| \delta \right| \gg
n_{o}U_{aa},n_{o}U_{bb},n_{o}U_{ab}.$ Consequently, condition (\ref
{splitmean}) should be experimentally achievable.

As mentioned before, in order for the speed of sound to show a dependence on
the internal states of the atoms, the scattering lengths $a_{ij}$ must be
different. For magnetically trapped $^{87}Rb$ with the two $5S_{1/2}$
hyperfine states $|a>=|F=1,m=-1>$ and $|b>=|F=2,m=1>,$ the scattering
lengths are nearly equal and are in the ratios $%
a_{aa}:a_{ab}:a_{bb}::1.03:1:0.97.$ This would preclude the use of $^{87}Rb.$
However, the recent demonstrations of the manipulation of the scattering
length in condensates of $^{85}Rb$ and $^{23}Na$ using a Feshbach resonance 
\cite{feshbach} opens up the possibility of using similar techniques in
multicomponent condensates. An alternative method for manipulating the
effective strength of the two-body interactions is based on a two-mode model
for the condensate in the presence of an external trapping potential \cite
{shift}\cite{twomode1}. This is discussed in appendix A.

\section{Summary and Conclusions}

We have shown that the atomic dressed states, which diagonalize the single
body Hamiltonian, are a useful basis for two-component condensates when the
dressed state energy splitting is much larger than the mean-field energies.
In this limit, the two-body interactions take on a simple form since those
interactions, in the dressed basis, which change the total number of atoms
in each of the two dressed states are highly suppressed. As such, the
Hamiltonian in the dressed basis has the same {\it form }as the Hamiltonian
in the original atomic basis with no external field coupling the two atomic
states. However, in the dressed basis, the coupling constants for the
two-body interactions are functions of the dressed state angle, a quantity
which is experimentally controllable. Consequently, collective properties of
the condensate which depend on the strength of the mean field interactions,
such as the speed of sound, may be controlled by adiabatically varying the
dressed state angle. The key requirement that $\hbar \omega _{cd}\gg
n_{o}U_{aa},n_{o}U_{bb},n_{o}U_{ab}$ represents a purely technical challenge
of achieving sufficiently high Rabi frequencies and detunings.

\section{Acknowledgments}

We are pleased to acknowledge helpful discussions with A. G. Rojo. This
research is supported by the National Science Foundation under grant No.
PHY-9800981 and by the U. S. Army Research Office under grant No.
DAAG55-97-0113 and No. DAAD19-00-1-0412.

\section{Appendix A- Anisotropic Trap.}

Up to now we have limited ourselves to the case of no external trapping
potential, $V_{i}({\bf r})\equiv 0$. However, another possibility for
manipulating the strength of the two-body interactions involves using
condensates in highly elongated cigar shaped traps. Let $Z$ be the length of
the condensate along the long axis, which we take to be in the z direction,
and $R$ the radius of the condensate. When the condition $kZ\gg 1$ and $%
kR\ll 1$ is satisfied, then the propagation of sound is along the long axis
of the trap and one may ignore changes in the radial direction. If one
ignores the trapping potential in the z-direction and considers a radial
harmonic potential with trapping frequencies $\omega _{a}$ and $\omega _{b}$
for the two-components, 
\begin{equation}
V_{i}(\rho )=\frac{1}{2}m\omega _{i}^{2}\rho ^{2}
\end{equation}
where $\rho =\sqrt{x^{2}+y^{2}}$ is the radial distance from the trap axis,
then one may express the field operators as 
\begin{mathletters}
\begin{eqnarray}
\hat{\Psi}_{a}({\bf r}) &=&\frac{1}{\sqrt{L}}\sum_{p}\alpha _{ap}\psi
_{a}(\rho )e^{ipz/\hbar };  \label{mode1} \\
\hat{\Psi}_{b}({\bf r}) &=&\frac{1}{\sqrt{L}}\sum_{p}\alpha _{bp}\psi
_{b}(\rho )e^{ipz/\hbar };  \label{mode2}
\end{eqnarray}
where $\psi _{i}(\rho )$ are the ground state radial wave functions for the
two states. Equations (\ref{mode1}-\ref{mode2}) represent a two mode
approximation with respect to the transverse coordinates. Physically, this
approximation consists of neglecting excitations above the ground state in
the directions transverse to the axis of the condensate.

When $\hbar \omega _{i}\gg n_{o}U_{kl}$, the $\psi _{i}(\rho )$ are the
harmonic oscillator ground states 
\end{mathletters}
\begin{mathletters}
\begin{equation}
\psi _{i}(\rho )=\left( \frac{m\omega _{i}}{\pi \hbar }\right)
^{1/2}e^{-m\omega _{i}\rho ^{2}/2\hbar }.  \label{ho}
\end{equation}
By substituting Eqs. (\ref{mode1}-\ref{mode2}) into Eqs. (\ref{hatom}-\ref
{hcoll}) and using Eq. (\ref{ho}), one obtains, after integration over the
spatial coordinates, an effective one-dimensional Hamiltonian for the
excitations in the axial direction which is given by, 
\end{mathletters}
\begin{mathletters}
\begin{eqnarray}
\hat{H}_{atom} &=&\sum_{p}\left\{ \left( \frac{p^{2}}{2m}+\frac{\hbar \delta 
}{2}+\hbar \omega _{a}\right) \alpha _{ap}^{\dagger }\alpha _{ap}+\left( 
\frac{p^{2}}{2m}-\frac{\hbar \delta }{2}+\hbar \omega _{b}\right) \alpha
_{bp}^{\dagger }\alpha _{bp}+\hbar \tilde{\Omega}_{R}\left( \alpha
_{bp}^{\dagger }\alpha _{ap}+\alpha _{ap}^{\dagger }\alpha _{bp}\right)
\right\} ;  \label{1dhatom} \\
\hat{H}_{coll} &=&\frac{1}{2L}\sum_{p_{1}+p_{2}=p_{3}+p_{4}}\left\{ \tilde{U}%
_{aa}\alpha _{ap_{1}}^{\dagger }\alpha _{ap_{2}}^{\dagger }\alpha
_{ap_{3}}\alpha _{ap_{4}}+\tilde{U}_{bb}\alpha _{bp_{1}}^{\dagger }\alpha
_{bp_{2}}^{\dagger }\alpha _{bp_{3}}\alpha _{bp_{4}}+2\tilde{U}_{ab}\alpha
_{ap_{1}}^{\dagger }\alpha _{bp_{2}}^{\dagger }\alpha _{ap_{3}}\alpha
_{bp_{4}}\right\} .  \label{1dhcoll}
\end{eqnarray}
One now has a rescaled Rabi frequency, $\tilde{\Omega}_{R}$, and one
dimensional mean-field interactions, $\tilde{U}_{ij}$ which are given by 
\end{mathletters}
\begin{mathletters}
\begin{eqnarray}
\tilde{\Omega}_{R} &=&\Omega _{R}\int 2\pi \rho d\rho \psi _{a}(\rho )\psi
_{b}(\rho )=\Omega _{R}\frac{2\sqrt{\omega _{a}\omega _{b}}}{\omega
_{a}+\omega _{b}}; \\
\tilde{U}_{aa} &=&U_{aa}\int 2\pi \rho d\rho |\psi _{a}(\rho )|^{4}=U_{aa}%
\frac{m\omega _{a}}{2\pi \hbar }; \\
\tilde{U}_{bb} &=&U_{bb}\int 2\pi \rho d\rho |\psi _{b}(\rho )|^{4}=U_{bb}%
\frac{m\omega _{b}}{2\pi \hbar }; \\
\tilde{U}_{ab} &=&U_{ab}\int 2\pi \rho d\rho |\psi _{a}(\rho )|^{2}|\psi
_{b}(\rho )|^{2}=U_{ab}\frac{m}{\pi \hbar }\frac{\omega _{a}\omega _{b}}{%
\omega _{a}+\omega _{b}}.
\end{eqnarray}
One sees that the Hamiltonian has the same form as Eq. (\ref{ha1}-\ref{ha2})
but for a one dimensional homogenous condensate. The harmonic oscillator
energies appearing in $\hat{H}_{atom}$ can be eliminated by a redefinition
of the internal energy levels for $\left| a\right\rangle $ and $\left|
b\right\rangle $. However, the one dimensional mean-field interactions, $%
\tilde{U}_{ij}$, can now be adjusted by varying the trap frequencies. For
example, if $\omega _{b}/\omega _{a}=2$ and $U_{ij}=U$ then $\tilde{U}_{bb}/%
\tilde{U}_{aa}=2$ and $\tilde{U}_{ab}/\tilde{U}_{aa}=4/3$. This case is
shown in Fig. 1(c).

The system described by Eqs. (\ref{1dhatom}-\ref{1dhcoll}) is different from
the Thomas-Fermi limit considered in \cite{soundtheory} \cite{stringari2}.
In the Thomas-Fermi limit, the radial wave function is a superposition of
harmonic oscillator excited states, $\psi _{i}^{(TF)}(\rho )=\sum_{m}c(\mu
,U_{aa},U_{bb},U_{ab})_{m}\psi _{i,m}(\rho ),$ where the $\psi _{i,m}(\rho )$
is the $m^{th\text{ }}$excited state of the radial harmonic oscillator and
the $c(\mu ,U_{aa},U_{bb},U_{ab})_{m}$ are functions of the chemical
potential and coupling constants(see Eq. (\ref{TF})). Consequently, the
resulting one dimensional form of $\hat{H}_{atom}$ will not be independent
of the total number of atoms (which is determined by $\mu $) or the strength
of the mean-field interactions in the Thomas-Fermi limit.

\end{mathletters}


\begin{references}
\bibitem{Na}  D.M. Stamper-Kurn {\it et al.,} Phys. Rev. Lett. {\bf 80},
2027 (1998).

\bibitem{Rb}  C. J. Myatt {\it et al.}, Phys. Rev. Lett. {\bf 78}, 586
(1997).

\bibitem{Na2}  Tin-Lun Ho, Phys. Rev. Lett. {\bf 81}, 742 (1998).

\bibitem{Na3}  T. Ohmi and K. Machida, J. Phys. Soc. Jpn. {\bf 67}, 1822
(1998).

\bibitem{Na4}  H. Pu, S. Raghavan, and N.P. Bigelow, Phys. Rev. A {\bf 61},
023602 (2000).

\bibitem{spin}  J. Stenger {\it et al.}, Nature {\bf 396, }345 (1998); D. M.
Stamper-Kurn, Phys. Rev. Lett. {\bf 83}, 661 (1999).

\bibitem{ramsey}  D. S. Hall, M. R. Mathews, C. E. Wieman, and E. A.
Cornell, Phys. Rev. Lett {\bf 81}, 1543 (1998).

\bibitem{Joseph}  J. Williams, R. Walser, J. Cooper, E. Cornell, and M.
Holland, Phys. Rev. A {\bf 59}, R31 (1999).

\bibitem{instable}  E. V. Goldstein and P. Meystre, Phys. Rev A {\bf 55},
2935 (1997).

\bibitem{shift}  E. V. Goldstein, M. G. Moore, H. Pu, and P. Meystre, Phys.
Rev. Lett. {\bf 85}, 5030 (2000).

\bibitem{dressed}  P.B. Blakie, R. J. Ballagh, and C.W. Gardiner, J. Opt. B:
Quantum Semiclass. Opt. {\bf 1}, 378 (1999).

\bibitem{collapse}  M. R. Mathews {\it et al.}, Phys. Rev. Lett {\bf 83, }%
3358 (1999); J. Williams {\it et al.}, Phys. Rev. A {\bf 61}, 033612 (2000).

\bibitem{dressedatom}  C. Cohen-Tannoudji, J Dupont-Roc, and G. Grynberg., 
{\it Atom-Photon Interactions} (Wiley, New York, 1992); P. R. Berman, Phys.
Rev. A {\bf 53}, 2627 (1996).

\bibitem{Gross}  L. P. Pitaevskii, Sov. Phys. JETP {\bf 13}, 451 (1961); E.
P. Gross, Nuovo Cimento {\bf 20}, 454 (1961); J. Math. Phys. {\bf 4}, 195
(1963).

\bibitem{stringari}  F. Dalfovo, S. Giorgini, L. P. Pitaevskii, S.
Stringari, Rev. Mod. Phys. {\bf 71}, 463 (1999).

\bibitem{bogo}  N. N. Bogoliubov, J. Phys. USSR {\bf 11}, 23 (1947).

\bibitem{sound}  M. R. Andrews {\it et al., }Phys. Rev. Lett. {\bf 79}, 553
(1997) and erratum, M. R. Andrews {\it et al., }Phys. Rev. Lett.{\bf 80},
2967 (1998).

\bibitem{soundtheory}  E. Zaremba, Phys. Rev. A {\bf 57}, 518 (1998); G. M.
Kavoulakis and C. J. Pethick, Phys. Rev. A {\bf 58}, 1563 (1998).

\bibitem{stringari2}  S. Stringari, Phys. Rev. A {\bf 58}, 2385 (1998).

\bibitem{two-photon}  In $^{87}Rb$, two-photon transitions involving a
microwave and rf field are used to couple the two hyperfine states. In this
case, $\omega _{e}$ would be the sum of the frequencies of the two fields
and $\Omega _{R}$ would be a two-photon Rabi frequency (see M. R. Mathews 
{\it et al.}, Phys. Rev. Lett. {\bf 81, }243 (1998)).

\bibitem{fetter}  A. L. Fetter and J. D. Walecka, {\it Quantum Theory of
Many-Particle Systems }(McGraw-Hill, New York, 1971).

\bibitem{rwa}  Marlan O. Scully and M. Suhail Zubairy, {\it Quantum Optics}
(Cambridge University Press, New York, 1997).

\bibitem{allen}  L. Allen and J. H. Eberly, {\it Optical Resonance and
Two-Level Atoms }(Dover, NY, 1987).

\bibitem{feshbach}  S. L. Cornish {\it et al.}, Phys. Rev. Lett. {\bf 85},
1795 (2000); S. Inouye {\it et al., }Nature (London) {\bf 392}, 151 (1998).

\bibitem{twomode1}  M. J. Steel and M. J. Collett, Phys. Rev. A {\bf 57},
2920 (1998).

\newpage

Fig 1. Plot of the speed of sound in units of $u=\sqrt{\frac{n_{o}U_{aa}}{m}}
$as a function of the dressed state angle $\theta .$ The dotted line is $%
u_{c}$ and the solid line is $u_{d}.$ The dressed sate angle is restricted
to the range $0\leq \theta \leq \pi /2.$ (a) $U_{bb}/U_{aa}=3/2$ and $%
U_{ab}/U_{aa}=2$ (b) $U_{bb}/U_{aa}=1/2$ and $U_{ab}/U_{aa}=2$ (c) $%
U_{bb}/U_{aa}=2$ and $U_{ab}/U_{aa}=4/3.$
\end{references}
\end{document}